\documentclass{PoS}

\title{Supporting the existence of the QCD critical point by compact star
observations}

\ShortTitle{CEP and compact star observations}

\author{\speaker{David E. Alvarez-Castillo}
	\\
       Bogoliubov Laboratory for Theoretical Physics, JINR Dubna, 141980 Dubna, Russia\\
       Instituto de F\'{i}sica, Universidad Aut\'{o}noma de San Luis Potos\'{i}, 78290 San Luis Potos\'{i}, 
       M\'{e}xico\\
       E-mail: \email{alvarez@theor.jinr.ru}}

\author{David Blaschke\\
	Bogoliubov Laboratory for Theoretical Physics, JINR Dubna, 141980 Dubna, Russia\\
	Instytut Fizyki Teoretycznej, Uniwersytet Wroclawski, 50-204 Wroclaw, Poland\\
        E-mail: \email{blaschke@ift.uni.wroc.pl}}

  \newcommand{\gt}{>}

\abstract{In order to prove the existence of a critical end point (CEP) in the
QCD phase diagram it is sufficient to demonstrate that at zero
temperature $T=0$ a first order phase transition exists as a  function
of the baryochemical potential $\mu$, since it is established
knowledge from ab-initio lattice QCD simulations that at $\mu=0$ the
transition on the temperature axis is a crossover.

We present the argument that the observation of a gap in the
mass-radius relationship for compact stars which proves the existence
of a so-called third family (aka "mass twins") will imply
that the $T=0$ equation of state of  compact star matter exhibits a
strong first order transition with a latent heat that satisfies
$\Delta\epsilon/\epsilon_c \gt 0.6$.

Since such a strong first order transition under compact
star conditions will remain first order when going to symmetric matter, the observation of a disconnected
branch (third family) of compact stars in the mass-radius diagram proves the existence of a CEP in QCD.
For the equation of state of the twins the quark matter description is based on a QCD-motivated chiral approach with higher-order quark interactions in the Dirac scalar and vector coupling channels. 
For hadronic matter we select a relativistic mean-field equation of state with density-dependent couplings. Since the nucleons are treated in the quasi-particle framework, 
an excluded volume has been included for the nuclear equation of state at super-saturation density that takes into account the finite size of the nucleons.

Furthermore we show results of a Bayesian analysis (BA) using disjunct M-R constraints for 
extracting probability measures for cold, dense matter equations of state. 
This study reveals that measuring the radii of neutron star twins has the potential to 
support the existence of a first order phase transition for compact star matter.}

\FullConference{9th International Workshop on Critical Point and Onset of Deconfinement - CPOD2014,\\
		17-21 November 2014\\
		ZiF (Center of Interdisciplinary Research), University of Bielefeld, Germany}

\begin{document}

\section{Introduction}

Neutron stars are compact objects which serve as astrophysical laboratories for studying extremely dense matter in their interiors \cite{Weber:1999qn,Glendenning:2000,Haensel:2007}, possibly composed of exotic forms of matter such as hypernuclear matter and deconfined quark matter \cite{Alford:2006vz}. 
Their investigation can be performed from both sides: astrophysical observations and nuclear collision experiments. 
Observational programs aim at measuring properties like mass, radius, moment of inertia, spinning and cooling rates among others. 
Nuclear experiments in the other hand look for precise determination of properties of nuclei in the laboratory via nuclear collisions and reactions.
In the last ten years important developments have occurred which allowed to place constraints to the equation of state (EoS) of neutron stars \cite{Klahn:2006ir}. 
These measurements are, however, not sufficient yet to completely determine the structure and composition of these compact stars and many questions remain open. 
The recent observation of the most massive neutron stars of about $2\,$M$_\odot$ 
\cite{Demorest:2010bx,Antoniadis:2013pzd}
has resulted in both ruling out EoS models that cannot support such massive stars and in restricting the range of densities in compact star matter.

Theoretical studies of dense matter have explored the QCD phase diagram and have shown that a smooth crossover transition occurs at very low chemical potentials and finite temperature - conditions for which colliders like the LHC and RHIC have provided compatible data. 
At higher chemical potentials and sufficiently low temperatures a first order phase transition to deconfined matter might occur which implies the existence of at least one critical end point (CEP) separating both phase transition types. 
The most massive neutron stars are likely to harbour deconfined quark matter in their cores 
 \cite{Alford:2006vz,Klahn:2006iw,Klahn:2013kga}
where values of the baryon density well exceed that of the nuclear saturation density and isospin asymmetry is extreme with proton fractions below 10\%. 
The existence of a CEP and onset of deconfinement (CPOD) results in important features in the neutron properties under certain conditions. 
The so called mass twin phenomenon being a most pronounced consequence of this
\cite{Alvarez-Castillo:2013cxa,Blaschke:2013ana,Benic:2014jia}. 
It is therefore relevant to point out what those observables are and to assess the feasibility of their detection. 
In this contribution we present the recent theoretical developments in the study of massive neutron
stars related to the high-mass twin phenomenon which can provide evidence for the existence of the CEP in the QCD phase diagram.

\section{The neutron star twins}

A strong first order phase transition to deconfined matter in neutron star matter will result in a disconnected branch in the mass-radius relation \cite{Alford:2013aca}
often called the third family after white dwarfs, and neutron stars
have been introduced as first and second families, respectively. 
The compact objects on the third family branch have a similar mass but smaller radii than pure neutron stars with a difference of up to two kilometers for the models discussed here.
This property led to the name "mass twins". 
While being known in the literature for a long time
\cite{Gerlach:1968zz,Kampfer:1981yr,Schertler:2000xq,Glendenning:1998ag,Dexheimer:2014pea}, the possibility of their existence at the high mass of $2\,$M$_\odot$ was demonstrated only recently 
\cite{Alvarez-Castillo:2013cxa,Blaschke:2013ana,Benic:2014jia}. 

A new microscopic approach for the description of hybrid stars was recently introduced in \cite{Benic:2014jia} incorporates important ingredients for the neutron star EoS and can also provide the mass twins. By introducing the effect of excluded volume of baryons
due to the Pauli exclusion principle between the quarks contained inside each nucleon the hadronic EoS stiffens up. On the other hand multi-quark interactions in the quark matter description also stiffen the EoS
at high densities  \cite{Benic:2014iaa}.

This results in a hybrid EoS for dense quark-baryon matter which supports massive hybrid stars of more than $2\,$M$_\odot$, eventually as a third family of compact stars. 
That case is equivalent to the existence of high-mass twins, with considerable radius differences.

\subsection{Why do we care about the twins?}

The suggestion to prove the presence of a CEP in the QCD phase diagram by the possible existence in nature of high-mass neutron star twins is in principle testable by astrophysical observations.
Besides this striking aspect of high-mass twins, their existence could solve also the presently most
urgent problems of compact star structure \cite{Blaschke:2015uva}:
\begin{itemize}
 \item {\bf The masquerade problem.} The EoS of hybrid stars with a smooth phase transition can look almost the same as the pure hadronic one. This results in a degenerate mass-radius relation which \textit{masquerades} hybrid stars as hadronic ones \cite{Alford:2004pf}.
 \item {\bf The hyperon puzzle.} The appearance of hyperons in neutron star interiors softens the EoS limiting the maximum stable star mass configuration that in most of the cases is much below the detected $2\,$M$_\odot$ value \cite{Baldo:2003vx}.
 \item {\bf The reconfinement puzzle.} In certain EoS a deconfinement transition occurs at lower densities but at much higher values the hadronic EoS becomes favourable again~\cite{Lastowiecki:2011hh,Zdunik:2012dj}.
 \item {\bf The measured radius issue.} The radius determination by measurements of thermal spectra from quiescent low-mass X-ray binaries inside globular clusters in ~\cite{Guillot:2013wu} is one of the lowest values reported. On the contrary,
 both measurements of hot spots in ~\cite{Bogdanov:2012md} and phase-resolved spectroscopy studies~\cite{Hambaryan:2014via} report rather large radius values. The nature of massive neutron stars twins feature favor long radii providing a resolution to the radius issue.
\end{itemize}

The neutron star twins features are rather robust against variations of the EoS, e.g., due to consideration of structures in the coexistence phase of the first-order phase transition ("pasta phases"). 
Although the appearance of pasta phases in the hadron-quark
interface will have an effect on the mass-radius diagram it will remove the third branch only for strong effects as explored phenomenologically in ~\cite{Alvarez-Castillo:2014dva}.
The third branch will persist also for fastly rotating stars as shown in ~\cite{Zdunik:2005kh} where stability
against radial oscillations was investigated. 
Studies for the presently discussed high-mass twin hybrid star EoS show that fast conversion scenarios 
from neutron stars to hybrid stars, e.g., by mass accretion are possible also under fast rotation for objects 
like the high-mass millisecond pulsar PSR J1614-2230 \cite{Demorest:2010bx} with a period of $3.15$ ms.

\section{Bayesian Analysis for hybrid neutron stars}

Bayesian analysis is a powerful statistical tool that can provide probabilities of detection of the hybrid twins phenomenon. 
In the context of constraining the EoS with observational data for masses and radii of compact stars it has first been introduced by \cite{Steiner:2010fz,Steiner:2012xt}. 
However their studies using X-ray bursters suffer from uncertainties related to the determination of the stellar atmospheres which is a crucial ingredient in spectral studies. 
Moreover, the parabolic region in the M-R diagram which follows from the determination of the 
luminosity radius $R_\infty$ is largely degenerate with an EoS of the APR type and thus fakes a high probability for such an EoS.
In \cite{Alvarez-Castillo:2014xea,Alvarez-Castillo:2014nua} the authors take a different approach by using a flexible parameterization of the EoS capable of including twin configurations on the one hand and is based on different neutron star measurements on the other, not relying on the atmosphere composition and being disjunct in mass and radius determination.

\section{Future studies on the twins}

There is still space left for exploration concerning the neutron star twins. From the stellar evolution side the open question of the feasibility of a dynamical transition from a hadronic neutron star into the hybrid one via accretion.
As mentioned before, rotation of hybrid stars has already been explored in \cite{Zdunik:2005kh} therefore predictions for the highly rotating massive neutron stars are in order. 
The physical imprints of gravitational radiation from excited modes is also a matter of investigation. Previous studies for hybrid stars going into this directions can be found in \cite{Nayyar:2007, Knippel:2009st}.

A future precise, simultaneous measurement of both a short and a large radius value for two objects is a crucial element for twin identification. Fictitious inclusion of such a measurement into the Bayesian analysis studies shall provide more realistic probabilities for the twin phenomenon appearance.

\section{Conclusions}

In this contribution we have reviewed the mass twin star phenomenon as a possible solution to several problems in modern astrophysics: the masquerade problem, the hyperon and reconfinement puzzles and the neutron star radius issue. 
Moreover, the observation of high-mass twins in nature would prove the existence of a critical point in the QCD phase diagram that is currently being looked for in terrestrial experiments with heavy ion collision experiments.  Since simulations of lattice QCD at finite temperature and vanishing baryon densities find a crossover behaviour, the observation of high-mass twin stars implying a strong first order phase transition at zero temperature and finite baryon density would be strong support for the existence of at least one critical endpoint in the QCD phase diagram. 
We have also pointed out that Bayesian analysis is a tool capable of assessing the feasibility of high-mass twin detection. 
In concluding we express our hope that future observational programmes with satellite missions like NICER \cite{nicer} or terrestrial large scale observatories like SKA \cite{ska} will provide the necessary accurate simultaneous determinations of masses and radii for a sufficiently large set of neutron stars which possibly will find support for the high-mass twin star phenomenon and thus the QCD critical endpoint.

\subsection*{Acknowledgements}
D.E.A-C. acknowledges support by the programmes for exchange between JINR Dubna and German Institutes (Heisenberg-Landau programme) as well as Polish Institutes (Bogoliubov-Infeld programme). This work was supported in part by the Polish National Science Center (NCN) under grant number 
UMO-2011/02/A/ST2/00306. 
The authors gratefully acknowledge the COST Action MP1304 "NewCompStar" for supporting their networking and collaboration activities.


\begin{thebibliography}{99}

\bibitem{Weber:1999qn} 
  F.~Weber,
  \emph{Pulsars as astrophysical laboratories for nuclear and particle physics},
  IOP, Bristol, 1999.

\bibitem{Glendenning:2000}
N.~K.~Glendenning, \emph{Compact Stars: Nuclear Physics, Particle Physics, and
General Relativity}, Springer, New York, 2000.

\bibitem{Haensel:2007}
P.~Haensel, A.~Y.~Potekhin and D.~G.~Yakovlev,
\emph{Neutron Stars I. Equation of State and Structure},
Springer, New York, 2007.

\bibitem{Alford:2006vz} 
  M.~Alford, D.~Blaschke, A.~Drago, T.~Kl\"ahn, G.~Pagliara and J.~Schaffner-Bielich,
  Nature {\bf 445}, E7 (2007).

\bibitem{Klahn:2006ir} 
  T.~Kl\"ahn, D.~Blaschke, S.~Typel, E.~N.~E.~van Dalen, A.~Faessler, C.~Fuchs, T.~Gaitanos, H.~Grigorian, A.~Ho, E.~E.~Kolomeitsev, M.~C.~Miller, G.~R\"opke, J.~Tr\"umper, D.~N.~Voskresensky, F.~Weber and H.~H.~Wolter, 
  Phys.\ Rev.\ C {\bf 74}, 035802 (2006).

\bibitem{Demorest:2010bx} 
  P.~Demorest, T.~Pennucci, S.~Ransom, M.~Roberts and J.~Hessels,
  Nature {\bf 467}, 1081 (2010)

\bibitem{Antoniadis:2013pzd} 
  J.~Antoniadis, P.~C.~C.~Freire, N.~Wex, T.~M.~Tauris, R.~S.~Lynch, M.~H.~van Kerkwijk, M.~Kramer, C.~Bassa, V.~S.~Dhillon, T.~Driebe, J.~W.~T.~Hessels, V.~M.~Kaspi, V.~I.~Kondratiev, N.~Langer, T.~R.~Marsh, M.~A.~McLaughlin, T.~T.~Pennucci, S.~M.~Ransom, I.~H.~Stairs, J.~van~Leeuwen, J.~P.~W.~Verbiest and D.~G.~Whelan,
  Science {\bf 340}, 6131 (2013).

\bibitem{Klahn:2006iw} 
  T.~Kl\"ahn, D.~Blaschke, F.~Sandin, C.~Fuchs, A.~Faessler, H.~Grigorian, G.~R\"opke and J.~Tr\"umper,
  Phys.\ Lett.\ B {\bf 654}, 170 (2007).

\bibitem{Klahn:2013kga} 
  T.~Kl\"ahn, R.~Lastowiecki and D.~B.~Blaschke,
  Phys.\ Rev.\ D {\bf 88}, no. 8, 085001 (2013).

\bibitem{Alvarez-Castillo:2013cxa} 
  D.~E.~Alvarez-Castillo and D.~Blaschke,
  arXiv:1304.7758 [astro-ph.HE].

\bibitem{Blaschke:2013ana} 
  D.~Blaschke, D.~E.~Alvarez-Castillo and S.~Benic,
  PoS CPOD {\bf 2013}, 063 (2013).

\bibitem{Benic:2014jia} 
  S.~Benic, D.~Blaschke, D.~E.~Alvarez-Castillo, T.~Fischer and S.~Typel,
  arXiv:1411.2856 [astro-ph.HE].

\bibitem{Alford:2013aca} 
  M.~G.~Alford, S.~Han and M.~Prakash,
  Phys.\ Rev.\ D {\bf 88}, no. 8, 083013 (2013).

\bibitem{Gerlach:1968zz} 
  U.~H.~Gerlach,
  Phys.\ Rev.\  {\bf 172}, 1325 (1968).

\bibitem{Kampfer:1981yr} 
  B.~K\"ampfer,
  J.\ Phys.\ A {\bf 14}, L471 (1981).

\bibitem{Schertler:2000xq} 
  K.~Schertler, C.~Greiner, J.~Schaffner-Bielich and M.~H.~Thoma,
  Nucl.\ Phys.\ A {\bf 677}, 463 (2000).

\bibitem{Glendenning:1998ag} 
  N.~K.~Glendenning and C.~Kettner,
  Astron.\ Astrophys.\  {\bf 353}, L9 (2000).
  
\bibitem{Dexheimer:2014pea} 
  V.~Dexheimer, R.~Negreiros and S.~Schramm,
  arXiv:1411.4623 [astro-ph.HE].

\bibitem{Benic:2014iaa} 
  S.~Benic,
  Eur.\ Phys.\ J.\ A {\bf 50}, 111 (2014).

\bibitem{Blaschke:2015uva} 
  D.~Blaschke and D.~E.~Alvarez-Castillo,
  arXiv:1503.03834 [astro-ph.HE].

\bibitem{Alford:2004pf}
  M.~Alford, M.~Braby, M.~W.~Paris and S.~Reddy,
  Astrophys.\ J.\  {\bf 629}, 969 (2005).

\bibitem{Baldo:2003vx} 
  M.~Baldo, G.~F.~Burgio and H.-J.~Schulze,
  in: \emph{Superdense QCD Matter and Compact Stars},
 edited by D.   Blaschke and D. Sedrakian, 
 Springer, Heidelberg, 2006, p.113; 
 [astro-ph/0312446].

\bibitem{Lastowiecki:2011hh} 
  R.~Lastowiecki, D.~Blaschke, H.~Grigorian and S.~Typel,
  Acta Phys.\ Polon.\ Supp.\  {\bf 5}, 535 (2012).

\bibitem{Zdunik:2012dj} 
  J.~L.~Zdunik and P.~Haensel,
  Astron.\ Astrophys.\  {\bf 551}, A61 (2013).
  
\bibitem{Guillot:2013wu} 
  S.~Guillot, M.~Servillat, N.~A.~Webb and R.~E.~Rutledge,
  Astrophys.\ J.\  {\bf 772}, 7 (2013).
  
\bibitem{Bogdanov:2012md} 
  S.~Bogdanov,
  Astrophys.\ J.\  {\bf 762}, 96 (2013).

\bibitem{Hambaryan:2014via} 
V.~Hambaryan, R.~Neuh{\"a}user, V.~Suleimanov and K.~Werner, 
J. Phys. Conf. Ser. {\bf 496}, 012015 (2014)


\bibitem{Alvarez-Castillo:2014dva} 
  D.~E.~Alvarez-Castillo and D.~Blaschke,
  arXiv:1412.8463 [astro-ph.HE].

\bibitem{Zdunik:2005kh} 
  J.~L.~Zdunik, M.~Bejger, P.~Haensel and E.~Gourgoulhon,
  Astron.\ Astrophys.\  {\bf 450}, 747 (2006).

\bibitem{Steiner:2010fz} 
  A.~W.~Steiner, J.~M.~Lattimer and E.~F.~Brown,
  Astrophys.\ J.\  {\bf 722}, 33 (2010).

\bibitem{Steiner:2012xt} 
  A.~W.~Steiner, J.~M.~Lattimer and E.~F.~Brown,
  Astrophys.\ J.\  {\bf 765}, L5 (2013).


\bibitem{Alvarez-Castillo:2014xea} 
  D.~Alvarez-Castillo, A.~Ayriyan, D.~Blaschke and H.~Grigorian,
  LIT Scientific Report 2011-2013, JINR Publishing Department, Dubna
  (2014) pp. 123-126. Editors: Gh. Adam, V.V. Korenkov, D.V. Podgainy, T.A.
  Strizh, P.V. Zrelov. ISBN 978-5-9530-0381-0
  [arXiv:1408.4449 [astro-ph.HE]].

\bibitem{Alvarez-Castillo:2014nua} 
  A.~Ayriyan, D.~E.~Alvarez-Castillo, D.~Blaschke, H.~Grigorian and M.~Sokolowski,
  arXiv:1412.8226 [astro-ph.HE].

\bibitem{Nayyar:2007} 
M.~Nayyar, Ph.D.~Thesis (2007).

\bibitem{Knippel:2009st} 
  B.~Knippel and A.~Sedrakian,
  Phys.\ Rev.\ D {\bf 79}, 083007 (2009).

\bibitem{nicer}
{\tt https://heasarc.gsfc.nasa.gov/docs/nicer}

\bibitem{ska}
{\tt http://www.ska.ac.za}

\end{thebibliography}
\end{document}